# On the Performance of FSO Communications Links under Sandstorm Conditions

Zabih Ghassemlooy[*], Joaquin Perez[*] and Erich Leitgeb[‡]

[*] Optical Communications Research Group, Faculty of Engineering and Environment,
Northumbria University, NE1 8ST, Newcastle Upon Tyne, United Kingdom
Email: {z.ghassemlooy, joaquin.perez} @northumbria.ac.uk

[‡] Institute of Microwave and Photonic Engineering,
Graz University of Technology, Inffeldgasse 12, Graz, Austria
Email: erich.leitgeb @tugraz.at

*Abstract*— In this paper we focus in the analysis of the FSO link performance under sandstorms conditions. The sandstorms are characterized by the size of the particles and the necessary wind speed in order to blowing them up during a minimum period of time. Sandstorm is a well know problem in many parts of the world and different reports has been presented regarding the impairments that sandstorms produces on outdoor link communications. The paper first focuses on the indoor laboratory sandstorms chamber and that is being used to investigate the performance on an FSO link. We propose an improvement to chamber with a dedicated customised structure where wind speed, sand blowing and turbulence can be generated, controlled and maintained over a much longer time period. This study would help in the deployment of a stable state of the art sandstorm environment for assessing FSO communication links.

I. INTRODUCTION

In recent years, the demand for FSO communications has increased considerably due to offering a large capacity usage (data, voice, and video) in a number of short to medium link range applications [1]. FSO offers a wider modulation bandwidth and aggregate data rate over 100 Gb/s using wavelength division multiplexing (WDM) [2]. It consumes a low power, provides much improved security against electromagnetic interference (EMI) and does not require administrative licensing or tariffs [3]. Therefore, FSO is considered as an alternative option in metropolitan and local area networks ((M/L)ANs) as well as a viable solution to the access network "last mile" problem in which the compromise between available data rates and the cost is desirable [4]. As it is well established and studied FSO communications face problems with aerosols particles in terms of atmospheric scintillation, turbulence and attenuation due to the interaction of this particles with the link propagation path [5-7]. Moreover, there is an additional problem due to sandstorm and dust storms as the result of increasing trend of desertification and urbanisation, which needs addressing particularly in arid and semi-arid regions [8]. Practical measurement of the weather impairments on outdoor FSO links has been reported based on long term observation of the link. However, there is very little study on the effects of dust storms and sandstorms, as part of a complex weather condition, on the FSO link performance. To be able to carry out a complete system measurement under all weather conditions, we have developed a dedicated indoor laboratory atmospheric chamber to investigate the effects of fog, smoke, temperature induced turbulence and windstorm and sandstorm and dust storm on the FSO link performance [9, 10]. In this paper a modification of the indoor chamber is proposed to replicate sandstorms and dust storm conditions. Regarding the results obtained a new laboratory structure to replicate sandstorm based on fluid dynamics simulations has been addressed and reported, following previous studies on the sandstorms dynamics [11, 12]. In this case we propose a new structure suitable for FSO communications links.

II. FSO LINK PERFORMANCES UNDER SANDSTORMS CONDITIONS

In this paper we demonstrate the feasibility study of sandstorm effects on an FSO link in a dedicated laboratory chamber. In the next sections the initial sandbox and measurement procedures are described. Then, the main results discussion, analysis and conclusion for the FSO link under the sandstorm for different modulation formats are summarized. Finally a new sandstorm structure is presented and discussed.

A. *Experimental indoor modified atmospheric chamber for sandstorm events*

Previously we have used facilities in the Optical Communications Research Group laboratory to study FSO communications links under atmospheric impairments as reported in [6, 7]. In this work we introduce another element onto the indoor atmospheric chamber in order to replicate sandstorm conditions. Previous study has shown that [13] the wind speed required to blow the sand depends on four major soil textures as outlines in Table I.

TABLE I. WIND VELOCITY (U) REQUIRED TO BLOW THE SAND REGARDING THE SOIL TEXTURE.

| Soil texture | U (m/s) |
|---|---|
| Dunes with thin sands | 5-8 |
| Sand with few pebbles | 10 |
| Flat desert | 10-13 |
| Desert with pebbles | 20 |

We have considered these wind speeds in order to modify the proposed indoor atmospheric chamber properly. Also, we have followed the standard classification for dust and sandstorms (DSS) climate events [14] that indicates dust and fine sand suspension (i.e. real sandstorm climate) normally occurs within a wind range of grades 4 to 11 (wind speed or U of 5.5-32.6 m/s). Then in this experiment, the range of wind speed is 7-16 m/s corresponding to a mid-level sandstorm scenario. We haven't considered wind scales higher than the grade 8 (wind speed of 17.0-20.7 m/s) that will result in mechanical malfunction of the FSO transceiver [15].

Therefore, we have proposed a new dedicated indoor atmospheric chamber adopted for the sandstorm study as shown in Figs. 1(a) and (b). The chamber has a dimension of 100×15×15 cm, with two fine filters at the top and bottom to allow full air circulation. Five fans positioned at different orientations are located at the base of the chamber each providing an air flow of ~16 m/s. In order to avoid the accumulation of fine sand in the vortex of the axial fan, five cone shape objects with the smooth surface are positioned between the fans as shown in Figs. 1(a) and (d). We have used fine sand with the average particle size of ~229 µm measured using the electron microscope image from environmental scanning electron microscope (ESEM), see Fig. 1(c). This sort of sand has a density of 1602 kg*m$^{-3}$, which corresponds to dry sand.

TABLE II. SET-UP PARAMETERS FOR THE FSO LINK UNDER STUDY

| Transmitter (Laser + AWG) | |
|---|---|
| Signal formats | OOK-NRZ, OOK-RZ and 4-PPM |
| Data-rate | 5 and 10 Mbit/s |
| Electrical Amplitude | {100, 250 and 500} mVpp |
| Operating wavelength | 785 nm |
| Average optical output power | -1.32 dBm @ 785 nm |
| Laser mod. depth | 20% |
| Laser 3-dB bandwidth | 50 MHz |
| Receiver (photodiode + lens) | |
| PD responsivity | 0.45 A/W @ 785 nm |
| PD active area | 5 mm2 |
| PD half angle view | ± 75º |
| Electrical filter at receiver | 0.75*|R| Hz |

An arbitrary waveform generator (AWG) is used to generate a stream of $2^{31}-1$ pseudo random bit sequence in OOK-NRZ, OOK-RZ and 4-PPM formats prior to the direct intensity modulation of a laser diode with a wavelength of 785 nm. The baseline data-rates chosen are 5 and 10 Mbit/s, and the optical power output measured is -1.32 dBm. At the receiver side, a photodiode followed by a wide-bandwidth transimpedance amplifier is employed to recover the electrical signal. The photodiode has a responsivity of 0.45 A/W at 785 nm. A digital storage oscilloscope is used to capture the signal for offline signal processing using Matlab software. A 1st order low pass filter with a cut-off frequency of 0.75*|R| Hz is used to reduce noise, where R = 1/T and T is the minimum pulse duration. The main system parameters are summarized in Table II.

In order to quantify the FSO link performance, the *Q*-factor is measured from the received signal using:

$$Q = \frac{(v_H - v_L)}{(\sigma_H - \sigma_L)} \quad (1)$$

where $v_H$ and $v_L$ are the mean received voltages and $\sigma_H$ and $\sigma_L$ are the standard deviations for the 'high' and 'low' level signals, respectively. Then BER is inferred from this expression:

$$BER = \frac{1}{2} * erfc\left(\frac{Q}{\sqrt{2}}\right) \quad (2)$$

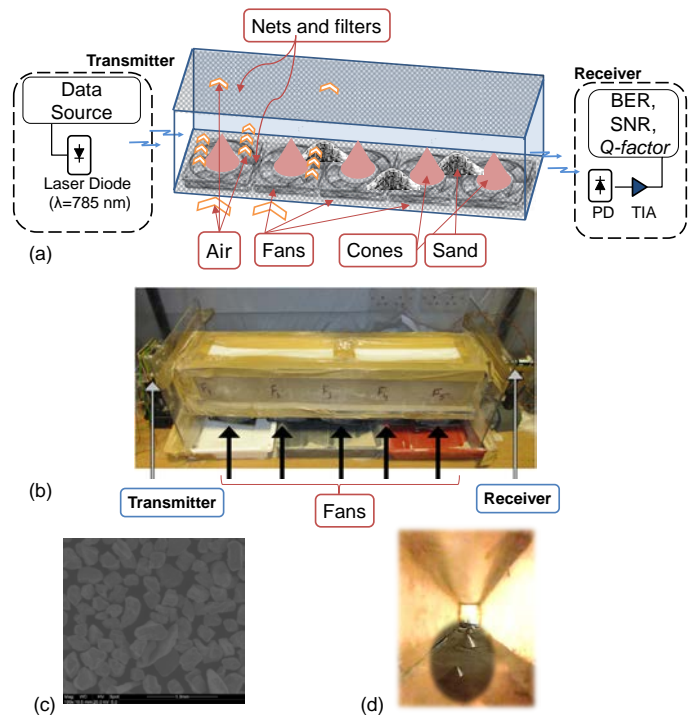

Fig. 1. (a) Experimental set-up block diagram to perform the sandstorm measurements. Snapshot of (b) experimental laboratory configuration of indoor from external view, (c) sand particles viewed at ESEM and (d) chamber from transmitter side internal view

### B. FSO communications link experimental results under sandstorm

The measured optical power in a clear condition with no sand is - 7.6 dBm dropping by more than 6 dB to - 13.8 dBm when there is a full wind speed sandstorm. We controlled the degree of sandstorm within the chamber with all the fans fully on, partially on and all off. In the case of fans fully on the attenuation measure at the receiver is defined by 0.24 transmittance factor. In order to compare with other atmospheric issues, e.g., fog, we calculate the correspondent visibility range using Kim's model for the laser beam propagation [10], obtaining a range below 70 m visibility range. According to the literature, we could classify the sandstorm as a secondary strong event, due to the low visibility and medium wind speed or velocity [14].

The effect of sandstorm on the FSO link performance is best depicted by the eye diagrams for clear and full sandstorm conditions for the OOK NRZ signal at 10 Mbit/s, as shown in Fig. 2.

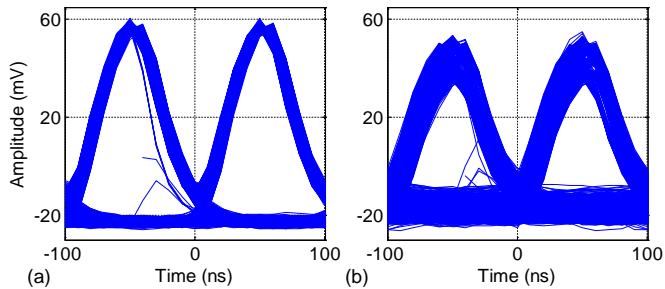

Fig. 2. Eye diagrams of the OOK–RZ signal received for 10 Mbit/s and 500 mVpp for a FSO link; (a) with no sandstorm and (b) with sandstorm

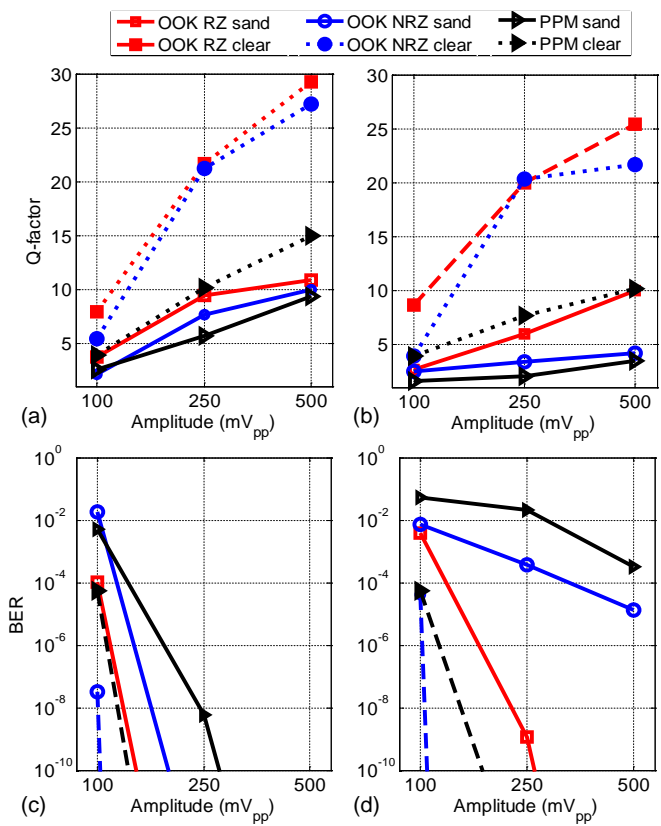

Fig. 3. Experimental results for the Q-factor and BER against the input signal amplitude for OOK-NRZ, OOK-RZ and 4-PPM at (a),(c) 5 Mbit/s and (b),(d) 10 Mbit/s with and without sandstorm, respectively.

From the received signal as depicted in Fig. 2 we measured the $Q$-factors for different modulation schemes for a 10 Mbit/s data rate with and without sandstorm. In Figs. 3(a) and (b) we show the measured $Q$-factor from the received signal for all data formats at 10 Mbit/s and 5 Mbit/s with and without sandstorm against different input signal amplitude. In the case of OOK-NRZ at 5 Mbit/s, for 100 mV$_{pp}$ the $Q$-factor falls from 7 to 4.5, however increasing the electrical signal amplitude results in $Q$-factor values of 7 and 9 for the sandstorms condition for 250 and 500 mV$_{pp}$, respectively. However for 10 Mbit/s with sandstorm OOK-NRZ $Q$-factor values are 2, 3 and 4.7 for 100, 250 and 500 mV$_{pp}$, respectively. In terms of the BER plot as a function of the signal amplitude are depicted in Figs. 3(c) and (d). For 10 Mbit/s data-rate the BER performance is much worse than the 5 Mbit/s for the entire range of the input signal.

From other point of view we decide to study if the wind speed associated with the sandstorm event also introduces turbulence effects on the FSO system. From the received signal it is possible to calculate the scintillation index $\sigma_I^2$ associated with the turbulence effect and related to the wind speed. This results obtained is always confined on the range of [0.2 0.4], thus indicating the turbulence is not a main degradation factor. Regarding $Q$-factor values obtained we checked that an improvement on signal amplitude does not implies a better scintillation index behaviour, as stated previously in [7].

### III. DESIGN OF A NEW SANDSTORM INDOOR ENVIRONMENT FOR FSO LINKS

After the evaluation of the experimental results obtained, explained in the previous section, the design of specialty indoor chamber to replicate the effects of sandstorm is explained in this section of the paper. Using the information on simulation of sandstorms and sand deposition [11, 12, 15] we have improved the sandstorm chamber using the fluid dynamics simulators. This enables us to incorporate sand particles dynamics in order to improve the replication of the sandstorm event in an indoor environment. The new study and design has been carried out using the ANSYS software and more specifically Fluent, which is used in Fluid Dynamics.

#### A. Fluid Dynamics and sandstorm chamber design

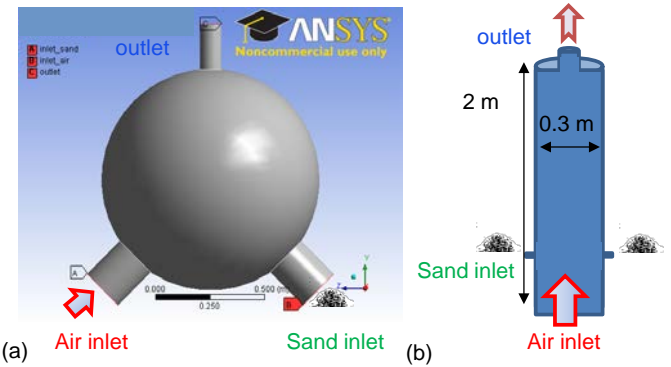

Fig. 4. Initial geometrical sandstorm designs to simulate in software based on (a) spherical and (b) cylindrical structure with inlets for sand, air and air outlet

First we considered spherical and cylindrical designs with pipes/apertures to introduce dust or sand particles into the chamber, see Fig. 4. Simulation results indicated that decreasing the diameter of the outlet enabled us to accelerate the particles exit, and so to prevent deposits of sand on walls. Comparing both models, the spherical model is not suitable for our design due to dimensions being larger than the cylindrical design in order to obtain a similar FSO propagation path to study the sandstorm effect. The cylindrical geometry offers a homogenous distribution of sand particles with two sand inlets in other sides of the tube. The main simulations due to the symmetry of the geometry selected, in this case cylindrical structure, were done on a 2D simulation environment in order to reduce the processing time on the software.

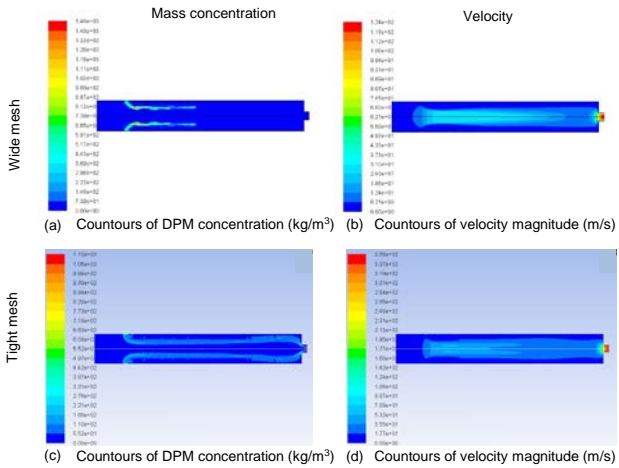

Fig. 5. Simulation results on Fluent (2D) with different meshes with mean velocity $v_{air} = 5$ m.s$^{-1}$; mass flow rate $q_m = 0.1$ kg.s$^{-1}$, (a)(c) different concentration of particles and (b)(d) for different air velocities.

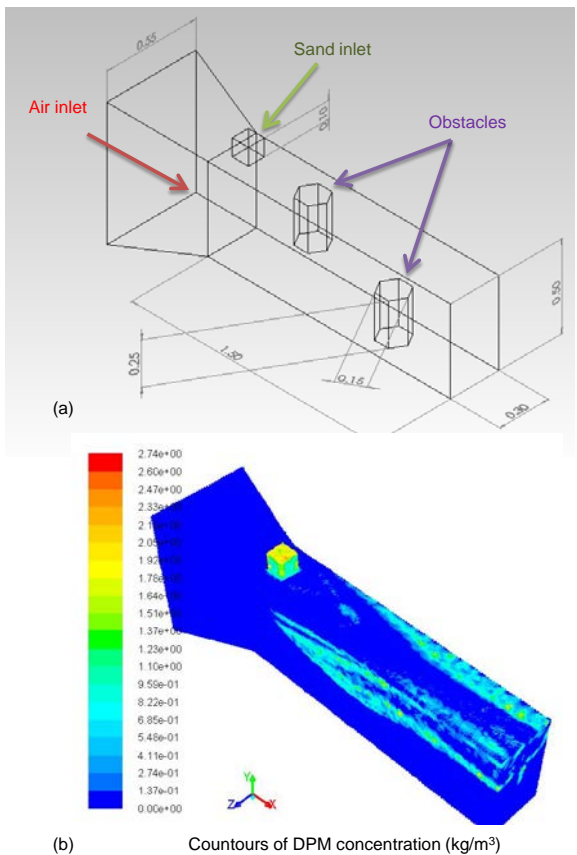

Fig. 6. (a) The design of obstacles on mechanical graphical design software and (b) the corresponding simulation on Fluent dynamics simulation software.

In simulation domain the definition of the mesh is very important in order to achieve the most accurate results. In this case particles are modelled as ash instead of sand, since sand particles are far bigger with a density of 1602 kg*m$^{-3}$, which corresponds to the dry sand. In this case the size of mesh should be approximately the size of the ash particle, which is usually between 20 and 500 µm. The initial simulations regarding the propagation of the sand particles and the velocity of each particle are depicted in Fig. 5. In both cases for different mesh size the air flow is maintained to $v_{air} = 5$ m/s. The injection of ash particles is at the some velocity as the wind speed. The different mesh particles indicates that a reduction on the sand inlet produces an acceleration on the particles but reduce their movement, as see in Fig. 5.

These simulations results led to a new idea of how to maintain the rate of sand inlet and avoid problems of sand inside distribution and deposition within the chamber.

In order to improve the turbulence environment within the chamber the area of interaction of sand and air with obstacles located at different locations was decreased, see Fig. 6(a). The simulation depicted result shown in Fig. 6(b) illustrate the generation of a more strong turbulence. Thus, indicating a approach to demonstrate the effect of sandstorm on the FSO link performance.

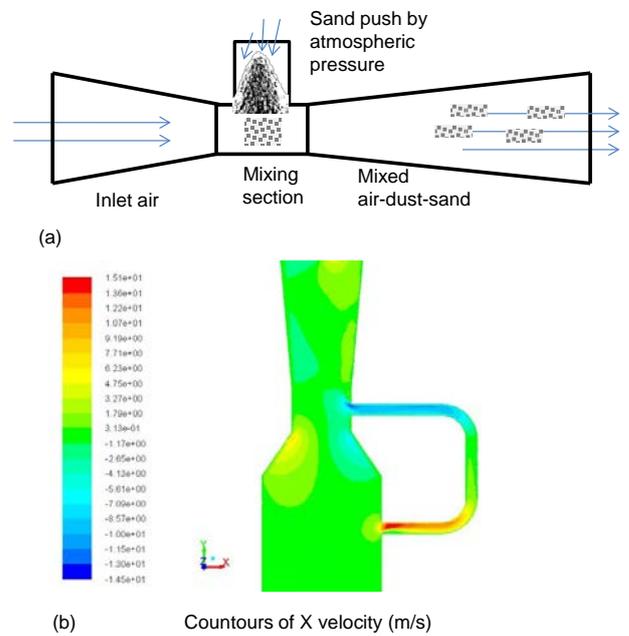

Fig. 7. (a) Typical Venturi effect structure on a tube, and (b) Simulation of the venture effect sand inlet design on the proposed sandstorm structure

A solution to answer the issue of injection of particles in the sandstorm chamber is the well-known Venturi effect by the way of introducing pressure differential to create an air flow and use it to blow particles inside the sandstorm machine, see Fig. 7(a). The Venturi effect in hydrodynamics describes the relation between the pressure of a non-viscosity fluid and the cross-section of the tubing it flows through, as a reduced cross-section leads to reduced pressure, as an example one of the first applications was the water jet pump proposed in 1869 by Bunsen [16] where the decrease of fluid pressure in a constriction helps to evacuate a side port. The velocity simulation in Fig 7(b) indicates the airflow and how it is possible to inject particles into the chamber.

Here we have chosen a square base where the laser and fans will be located and will generate a homogeneous wind throughout the structure, thus avoiding sand deposition in the corners. At the top of the pipe, the photodetector is located.

The final design regarding the previous concepts is focus on build a test device for FSO communication links. We select the minimum height of the chamber at 1500 mm and we located the sand injection via a valve regulated by a funnel in order to control the flow of particles. The outlet is modified to collect the sand particles using filters similar to a vacuum cleaner deposit. The chamber internal pressure could be up to 1.5 bars. The final proposed design regarding all these aspects is shown in Fig. 8.

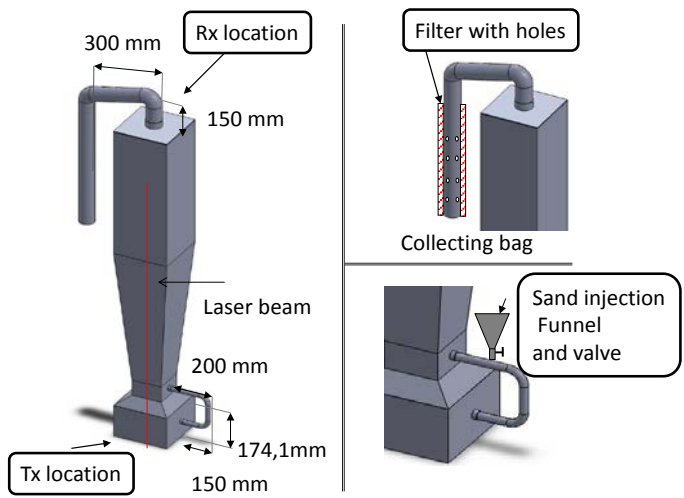

Fig. 8. Final sandstorm designed according to ANSYS Fluent simulation results, details of sand inlet and air outlet sand collection

### B. Experimental test on turbulence improvement due to inner obstacle

As it is shown in Fig. 9(a) we have used a small-scale wind tunnel with two nozzles, an exhaust fan which pump the air injected and a speed controller which determined the rotational speed of the fan. In order to observe the air spreading within the chamber with obstacles in places, we have injected fog, see Fig. 9(b). The experiment was carried for 5, 8 and 10 m/s wind speed. Fig. 10 shows the vortex observed after the obstacle and beyond the obstacles the fog flow decreased consistently, thus leading to increase of flow on either side of the obstacle.

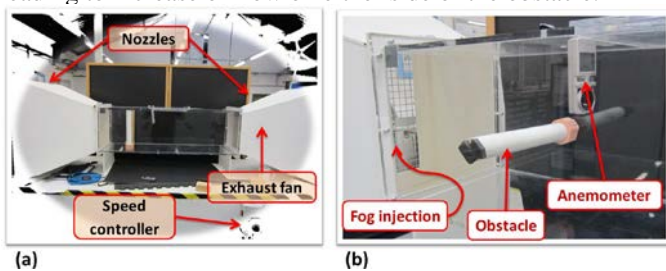

Fig. 9. Snapshot of the (a) wind tunnel and (b) laboratory set-up used to test turbulence due to obstacles

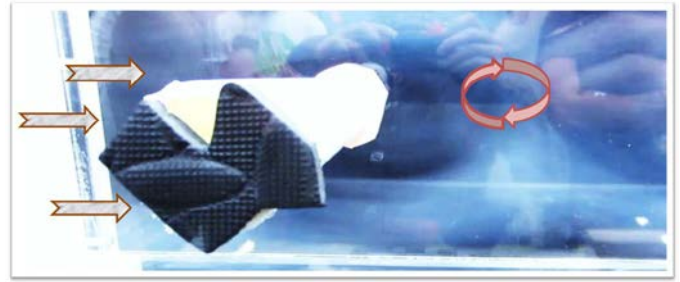

Fig. 10. Snapshot of the experimental results for a speed of fog 8 m.s$^{-1}$

## IV. CONCLUSION

In this paper we have demonstrated the feasibility of replicating sandstorms conditions under a controlled laboratory environment. This enabled us to test the FSO link performance and the link availability under sandstorms effects. Experimental results showed that the FSO link performance can be improved by increasing the transmitted signal amplitudes and selecting adequate line rates. Due to the size of the particles the behaviour of this phenomenon is comparable to the fog attenuation over a short time periods. We have reported a new dedicated sandstorm chamber to test FSO links.

### ACKNOWLEDGMENT

This research has been supported by the European COST action IC0802 "Propagation tools / data for integrated Telecom, Navigation and Earth Observation systems" and action IC1101 "OpticWise - Optical Wireless Communications - An Emerging Technology".

### REFERENCES

[1] E. Ciaramella, Y. Arimoto, G. Contestabile, M. Presi, A. D'Errico, V. Guarino, and M. Matsumoto, "1.28 terabit/s (32x40 Gbit/s) wdm transmission system for free space optical communications," *Selected Areas in Communications, IEEE Journal on,* vol. 27, pp. 1639-1645, 2009.
[2] N. Cvijetic, Q. Dayou, Y. Jianjun, H. Yue-Kai, and W. Ting, "100 Gb/s per-channel free-space optical transmission with coherent detection and MIMO processing," in *35th European Conference on Optical Communication ECOC* 2009, pp. 1-2.
[3] Z. Ghassemlooy, W. Popoola, and S. Rajbhandari, *Optical wireless communications : system and channel modelling with MATLAB*. Boca Raton, FL: CRC Press 2012.
[4] E. Leitgeb, M. Gebhart, and U. Birnbacher, "Optical networks, last mile access and applications," *Journal of Optical and Fiber Communications Research,* vol. 2, pp. 56-85, 2005.
[5] F. Nadeem, V. Kvicera, M. S. Awan, E. Leitgeb, S. Muhammad, and G. Kandus, "Weather effects on hybrid FSO/RF communication link," *Selected Areas in Communications, IEEE Journal on,* vol. 27, pp. 1687-1697, 2009.
[6] J. Perez, Z. Ghassemlooy, S. Rajbhandari, M. Ijaz, and H. L. Minh, "Ethernet FSO Communications Link Performance Study Under a Controlled Fog Environment," *IEEE Communications Letters,* vol. 16, pp. 408-410, 2012.
[7] Z. Ghassemlooy, H. Le Minh, S. Rajbhandari, J. Perez, and M. Ijaz, "Performance Analysis of Ethernet/Fast-Ethernet Free Space Optical Communications in a Controlled Weak Turbulence Condition," *Lightwave Technology, Journal of,* vol. 30, pp. 2188-2194, 2012.

[8] A. R. Raja, Q. J. Kagalwala, T. Landolsi, and M. El-Tarhuni, "Free-Space Optics Channel Characterization under UAE Weather Conditions," presented at the Signal Processing and Communications, 2007. ICSPC 2007. IEEE International Conference on, 2007.

[9] W. O. Popoola, Z. Ghassemlooy, C. G. Lee, and A. C. Boucouvalas, "Scintillation effect on intensity modulated laser communication systems - a laboratory demonstration," *Optics & Laser Technology,* vol. 42, pp. 682-692, 2009.

[10] J. Perez, Z. Ghassemlooy, S. Rajbhandari, M. Ijaz, and H. Minh, "Ethernet FSO Communications Link Performance Study Under a Controlled Fog Environment," *Communications Letters, IEEE,* vol. 16, pp. 408-410, 2012.

[11] N. Wang and B.-G. Hu, "Real-Time Simulation of Aeolian Sand Movement and Sand Ripple Evolution: A Method Based on the Physics of Blown Sand," *Journal of Computer Science and Technology,* vol. 27, pp. 135-146, 2012/01/01 2012.

[12] Z. Ming and J. Wang, "The study of the sand density distribution in the experimental segment of the sand wind tunnel by computer simulation," in *Future Computer and Communication (ICFCC), 2010 2nd International Conference on*, 2010, pp. V1-462-V1-466.

[13] W. Nickling and C. Neuman, "Aeolian Sediment Transport," in *Geomorphology of Desert Environments*, A. Parsons and A. Abrahams, Eds., ed: Springer Netherlands, 2009, pp. 517-555.

[14] V. R. Squires, "Physics, Mechanics and Processes of Dust and Sandstorms," in *GLOBAL ALARM: DUST AND SANDSTORMS FROM THE WORLD'S DRYLANDS*, Yang Youlin, Victor R. Squires, and L. Qi, Eds., ed Bangkok: Secretariat of the UNCCD, Asia RCU, 2001, pp. 15-77.

[15] H. Bo, J. Haiyun, G. Naikui, C. Bangfa, and P. Zongren, "Characteristics of dust deposition on suspended insulators during simulated sandstorm," *Dielectrics and Electrical Insulation, IEEE Transactions on,* vol. 17, pp. 100-105, 2010.

[16] R. Bunsen, "I. On the washing of precipitates," *Philosophical Magazine Series 4,* vol. 37, pp. 1-18, 1869/01/01 1869.